\documentclass[aps,pra,nofootinbib,preprint,
showpacs,amsmath,amssymb,floatfix,showpacs]{revtex4}
\usepackage{amssymb}
\usepackage{amsmath}
\usepackage{bm}
\usepackage{upgreek}
\usepackage{graphicx}

\begin{document}

\title{Analysis of recent interpretations of the Abraham-Minkowski problem}         

\author{Iver Brevik}
\address{Department of Energy and Process Engineering, Norwegian University of Science and Technology, 7491 Trondheim, Norway}

\date{\today}          

\begin{abstract}
The 100-year-long problem concerning the correct form of the electromagnetic energy-momentum tensor in continuous media (usually called the Abraham-Minkowski problem) continues to attract interest. Here we provide a  critical analysis of   interpretations presented in the literature on this topic recently, in two cases -  one real experiment on radiation pressure A. Kundu {\it et al.}, Scient. Reports {\bf 7}, 42538 (2017), and one computer experiment M. Partanen {\it et al.}, Phys. Rev. A {\bf 95}, 063850 (2017).

\end{abstract}

\maketitle

\section{Introduction} \label{sec1}

The long-standing discussion about what is the correct form of the electromagnetic energy-momentum tensor in a medium has recently become accentuated again. This problem - usually called the Abraham-Minkowski problem \cite{abraham09,minkowski10} - is to determine the correct expression for the electromagnetic momentum in the medium, or equivalently, whether there exists a special term called the "Abraham term" $\bf{f}^{\bf A}$ (see Eq.~(\ref{2}) below) in the electromagnetic force density $\bf f$.  We have recently considered the Abraham-Minkowski problem both classically and quantum mechanically, from various points of view, though  with a particular emphasis on radiation pressure phenomena \cite{brevik10,brevik17,brevik18}. There exists quite naturally a big numbers of other papers in this area also; some of them are listed in Refs.~\cite{ashkin73,astrath14,zhang15,verma15,schneider16,partanen17,nesterenko16,nesterenko16a,barnett15,bethunewaddell15,kemp15,conti14,kemp13,leonhardt14,barnett10,
baxter10,barnett10a,hinds09,mansuripur10,pfeifer07,loudon05,loudon02}.

The chief purpose of this note is to give an analysis of two statements put forward in the recent literature, both of them of significance in the Abraham-Minkowski context:

\bigskip

\noindent 1) The radiation pressure experiment of Kundu {\it et al.} \cite{kundu17} gave a clear demonstration of how a weak cw laser beam falling from above on a graphene oxide film deflects the film in the downward direction, typically by an amount of about 80 nm, when the laser power was about $P=1.4~$mW. The authors  interpreted the experiment so as to favor the Abraham expression for the photon momentum in matter. Is this interpretation right?

\noindent 2) Consider another situation: when an optical pulse propagates in an (infinite) isotropic medium, the Abraham term ${\bf f}^{\rm A}$ exerts a longitudinal force on the matter, in particular at the leading  and trailing edges of the pulse. In the paper of Partanen {\it et al.} \cite{partanen17} it was argued that this means  transfer of quite a large  mechanical energy  from the pulse to the medium,  of the same order of magnitude as the field energy itself. Again, is this  interpretation right?

\bigskip

 It is now necessary to write down some central formulas. For an isotropic nonmagnetic medium with permittivity $\varepsilon=n^2$ the Abraham force density can be expressed as
\begin{equation}
{\bf f}=\rho {\bf E}+ \mu_0{\bf J\times H}- \frac{1}{2}\varepsilon_0E^2 \nabla n^2 +\frac{n^2-1}{c^2}\frac{\partial}{\partial t}\bf (E\times H), \label{1}
\end{equation}
cf., for instance, Refs.~\cite{brevik79} or \cite{landau84}; we here  write the constitutive relations as ${\bf D}=\varepsilon_0\varepsilon {\bf E},\, {\bf B}=\mu_0{\bf H}$, and omit the electrostriction term.

In Eq.~(\ref{1}), $\rho$ and $\bf J$ are the macroscopic charge and current densities. When the medium is at rest as assumed here, $\bf J$ is a pure conduction current. Ohm's law is ${\bf J}=\sigma \bf E$, with $\sigma$ the electrical conductivity. When dealing with problems in optics, $\sigma$ and $\bf J$ are often left out. The third term in Eq.~(\ref{1}), proportional to the gradient of the permittivity,  is of importance at dielectric boundaries. It is common for the Abraham and Minkowski theories and may thus be called the Abraham-Minkowski term. It significance   has been demonstrated in several optical experiments.  (One may say that the development in this direction started with the classic experiment of Ashkin and Dziedzic in 1973 \cite{ashkin73}: a narrow light beam impinging vertically from above on a free water surface acted upon the surface by an outward pull. Using pulsed radiation of peak power $P=$3 kW, each pulse of duration 60 ns, they observed an elevation of the surface of about 0.9 $\mu$m. This experiment was analyzed also in Ref. \cite{brevik79}. Later experiments have shown surface defections of considerably higher magnitude; cf., for instance, the case where  two immiscible fluids are situated above each other when the temperature is near to the critical point  \cite{casner03,wunenburger11}.)

The fourth term in Eq.~(\ref{1}) is the mentioned Abraham term,
\begin{equation}
{\bf f}^{\rm A}=\frac{n^2-1}{c^2}\frac{\partial}{\partial t}{\bf (E\times H)}= \frac{n^2-1}{c^2}\frac{\partial {\bf S}}{\partial t}, \label{2}
\end{equation}
with $\bf S$ the Poynting vector. Under stationary conditions in optics  this term does not contribute at all; it  fluctuates out.

In the next three sections we will consider the mentioned items 1) and 2) separately, and supply with some considerations about the dynamics of a laser  illuminated  vibrating graphene sheet.

\section{Radiation pressure on a graphene oxide plate}

We consider the same basic setup as in  the experiment \cite{kundu17}, namely a horizontal graphene oxide plate illuminated by a weak cw laser beam from above, in the vertical $z$ direction. We take the wave to be   polarized in the $x$ direction so that  $E_x=E_0e^{i(k_0z-\omega t)}$ is the only nonvanishing component ($k_0=\omega/c$). The Poynting vector of the incident wave is thus
\begin{equation}
S_I=\frac{1}{2}\varepsilon_0 cE_0^2. \label{3}
\end{equation}
We let the $z$  axis point downwards, and let  the upper surface  of the plate lie at $z=0$; the lower surface lies at $z=d$, where $d$ is the thickness.

The mechanical force on the plate is composed by two different contributions:

\noindent (i) There are vertical forces acting at the boundaries $z=0$ and $z=d$, due to the gradient term (the third term in Eq.~(\ref{1})). At $z=0$ the force acts upwards, at $z=d$ it acts downwards ($n>1$ assumed). A specific calculation, not shown here,  assuming (unrealistically) the refractive index to be real, leads to a total gradient force pointing downwards.

\noindent (ii) There are vertical Lorentz forces acting in the interior of the plate because of the  conductivity $\sigma$ of graphene oxide. The refractive index of this material is complex; calling it $\tilde{n}$, we have as the mean value at wavelength 532 nm \cite{wang08}:
\begin{equation}
\tilde{n} =2.4 +1.0~i \label{4}
\end{equation}
(the plus sign occurs due to our convention $e^{-i\omega t}$). This means that  $\sigma$ is quite large. While  electrodynamic theory in metals is complicated \cite{stratton41},  we will henceforth as a first approximation ignore  the surface forces considered above and focus on  the Lorentz force only.

For large $\sigma$ the theory of metals can be simplified significantly. Formally, this corresponds to the limit  $k_0/\alpha \ll 1$, where $k_0$ is the incident wave number as before, and $\alpha$ is defined as
\begin{equation}
\alpha=\sqrt{\mu_0 \omega \sigma/2}. \label{5}
\end{equation}

As mentioned, we assume  that the wave falls normally upon the plate at  the surface $z=0$, the plate now for convenience taken to be infinitely thick. The approximate expressions for the fields in the two regions are:
\begin{equation}
E_x=E_0\left[ e^{i(k_0z-\omega t)}-
{\sqrt{R}}\,e^{-i\delta}e^{-i(k_0z+\omega t)}\right], \, (z<0), \label{6}
\end{equation}
\begin{equation}
H_y=\frac{k_0E_0}{\mu_0\omega}\left[ e^{i(k_0z-\omega t)}+
{\sqrt{R}}\,e^{-i\delta}e^{-i(k_0z+\omega t)}\right], \, (z<0), \label{7}
\end{equation}
\begin{equation}
E_x=\frac{k_0E_0}{\alpha}(1-i)e^{-\alpha z}e^{i(\alpha z-\omega t)}, \, (z>0), \label{8}
\end{equation}
\begin{equation}
H_y=\frac{k_0E_0}{\mu_0\omega}\left[2-(i-1)\frac{k_0}{\alpha}\right] e^{-\alpha x}e^{i(\alpha z-\omega t)}, \, (z>0), \label{9}
\end{equation}
with
\begin{equation}
R=1-2k_0/\alpha, \quad \tan \delta=-k_0/\alpha. \label{10}
\end{equation}
These expressions satisfy the boundary conditions at $z=0$ to the first order in $k_0/\alpha$.

We can now calculate the force on unit area of the plate by integrating the Lorentz force over the appropriate volume limited by $z=0$ and $z=d$,
\begin{equation}
\sigma_z=\frac{1}{2}\mu_0\sigma \Re \int_{0}^{d}E_xH_y^* dz. \label{11}
\end{equation}
Insertion of the above expressions gives
\begin{equation}
\sigma_z=\varepsilon_0E_0^2(1-e^{-2\alpha d})(1-\frac{k_0}{\alpha}). \label{12}
\end{equation}

With $\lambda_0=532~$nm, $\omega=3.54\times 10^{15}~$rad/s we get $\alpha=4.72\times 10^4\times \sqrt{\sigma}$, or $k_0/\alpha=250/\sqrt{\sigma}$. As an example, we may choose
\begin{equation}
\sigma=1.0\times 10^6~\rm{S/m}, \label{13}
\end{equation}
which is about the same  conductivity as for manganese steel. Then, $k_0/\alpha=0.25$, and the above condition is roughly  satisfied. With $d=300~$nm  the term $e^{-2\alpha d}\ll 1$, and we obtain
\begin{equation}
\sigma_z=\varepsilon_0E_0^2\times 0.75. \label{14}
\end{equation}

One may ask if our choice (\ref{13}) for the conductivity is reasonable. It corresponds to a two-dimensional sheet conductivity equal to $\sigma^{2D}=3~$Sm, when $d=300~$nm. This is a quantity that is in principle accessible experimentally.
To get some more insight at this point, let us go back to Eq.~(\ref{4}) for the complex refractive index and calculate the complex permittivity,
\begin{equation}
\varepsilon =\tilde{n}^2= (2.4+1.0\,i)^2=4.76+4.8\,i. \label{15}
\end{equation}
In conventional notation $\varepsilon=\varepsilon'+i\varepsilon''$; thus  $\varepsilon'=4.76,\, \varepsilon''=4.8$. Now comparing with the formula
\begin{equation}
\varepsilon=\varepsilon'+\frac{i\sigma}{\varepsilon_0\omega}, \label{16}
\end{equation}
we obtain the value $\sigma=1.5\times 10^5~$S/m. Although there are considerable uncertainties when associating two-dimensional sheets with three-dimensional quantities,   this indicates that our calculation  has overestimated  $\sigma$ a  bit. We might have used a lower value of $\sigma$ (giving a weaker surface force), but at the expense of violating the condition $k_0/\alpha \ll 1$.

We will not enter into further detail here, but conclude that the expression (\ref{14}) should give a reasonable value for the Lorentz force on the plate. When augmented by the surface force at the boundaries (not shown, as mentioned) we actually get a value for $\sigma_z$  that becomes roughly the same as for  total reflection,
\begin{equation}
\sigma_z=\varepsilon_0E_0^2. \label{17}
\end{equation}
We will for definiteness use this expression as the driving force in the following. It is in agreement with the assumption made in Ref.~\cite{kundu17}.

\section{Statics and dynamics of the circular plate}

We will apply  elasticity theory to estimate the influence from the force (\ref{17}) on the graphene oxide plate. Now, a practical complication in the experiment \cite{kundu17}  was that the plate was residing on a Si substrate. It is natural to  assume that there was not a direct mechanical contact between plate and substrate; otherwise there would be no deflection at all. Moreover, if transmission properties in the plate were allowed for, it would be necessary to include the optical properties of the substrate also. We will henceforth avoid these possible complications by assuming that the plate is surrounded by air ($n=1$) both on the upper and the lower side. Such a simplified  model is yet able to demonstrate the essence of the effect.

Thus take the plate of thickness $d$ to be circular, of radius $a$, and assume for simplicity that the pressure  $\sigma_z$ is constant over the initially flat cross section $\pi a^2$.  The total radiation force is thus
\begin{equation}
F_z=\frac{2P}{c}, \label{18}
\end{equation}
where the incident power is
\begin{equation}
P=S_I\pi a^2=\frac{1}{2}\varepsilon_0cE_0^2\times \pi a^2. \label{19}
\end{equation}

\subsection{Statics}

We will first evaluate the form of the plate in its equilibrium state when acted upon by the cw laser beam.
 Adopt cylindrical coordinates with the origin lying at the center of the undisturbed sheet and let, as mentioned,  the $z$ axis be pointing downwards. The stationary deflection  $\zeta$ depends on the radius, $\zeta=\zeta(r)$. The governing equation for large deflections   is in general quite complicated,   of the fourth order in $\zeta$ \cite{landau86}.

 We will model the graphene sheet as an  elastic plate subject to the conditions that both the elevation and the slope of the plate are zero at  $r=a$ (i.e., a clamped edge situation).  The governing equation is
\begin{equation}
D\nabla^4 \zeta=\sigma_z +\rho g d, \label{20}
\end{equation}
where
\begin{equation}
D= \frac{Ed^3}{12(1-\nu^2)} \label{21}
\end{equation}
is the flexural rigidity. Here  $E$ is Young's modulus,  and $\nu$ is Poisson's ratio. For graphene,  $\nu \approx 0.16$ \cite{lee08}.
Equation (\ref{20}) is quite general,  holding even if the deflection $\zeta$ is large compared with  $d$.
With $a=0.9 {\mu}$m, $P=1.4$ mW we find $\sigma_z=3.67~$Pa, while the gravitational pressure is much less, $\rho gd=6.65 $ mPa, when $\rho=2.26~$g/cm$^3$. Thus the term $\rho gd$ can be omitted,  and   the equation reduces to
\begin{equation}
\frac{1}{r}\frac{d}{dr}\left\{ r\frac{d}{dr}\left[ \frac{1}{r}\frac{d}{dr}\left( r\frac{d\zeta}{dr}\right)\right]\right\} = \frac{\sigma_z}{D},  \label{22}
\end{equation}
which by integration yields
\begin{equation}
\zeta=\frac{\sigma_za^4}{64D}\left( 1-\frac{r^2}{a^2}\right)^2.
\end{equation}
The slope at $r=a$ is thus zero.
 Using the maximum deflection  from the experiment, $\zeta_{\rm max}=\sigma_za^4/(64D)=80$ nm, we can  estimate the effective value of the flexural rigidity. We find $D=4.7\times 10^{-19}$ Nm, corresponding to a very low value
\begin{equation}
E=200~{\rm Pa}. \label{23}
\end{equation}

\subsection{Dynamics}

 The dynamic aspects of the problem are also of interest, although this is a topic  not directly connected with the experiment  \cite{kundu17}. We will confine us to the case of free vibrations, i.e., put $\sigma_z=0$.

  Assume still that the graphene oxide sheet is modeled as a circular plate of thickness $d$ clamped along its periphery $r=a$. The governing equation is
  \begin{equation}
  D\nabla^4 \zeta + \rho d \ddot{\zeta}=0.
  \end{equation}
  With the basic ansatz $\zeta(r,t)=W(r)\cos \omega t$ we get
  \begin{equation}
  \nabla^4 W=\lambda^4 W,
  \end{equation}
  with
  \begin{equation}
  \lambda^4=\frac{\rho d \omega ^2}{D}.
  \end{equation}
  The general solution for radially symmetric deflections can be expressed in terms of ordinary and modified  Bessel functions,
  \begin{equation}
  \zeta r,t)=\sum_{n=1}^\infty C_n\left[ J_0(\lambda_n r)-\frac{J_0(\lambda_na)}{I_0(\lambda_na)}I_0(\lambda_nr)\right] \cos \omega_n t.
  \end{equation}
 The boundary conditions for a clamped plate are therewith satisfied: $W(r)=0$ and $dW/dr=0$ at $r=a$.

 As for the eigenfrequencies, it is convenient to make use of the approximative formula given by Timoshenko \cite{timoshenko55},
 \begin{equation}
 \omega =\frac{\alpha}{a^2}\sqrt{ \frac{D}{\rho d}}, \label{A}
 \end{equation}
 where $\alpha$ is a constant characteristic for the mode. We will consider only the lowest mode, for which $\alpha=10.21$. Then $\omega$ follows from Eq.~(\ref{A}), whereas $\lambda$ is conveniently found as $\lambda= \sqrt{\alpha}/a$.

 Inserting $\alpha=10.21,\, a=0.9~\mu$m, $D=4.7\times 10^{-19}~$Nm, $\rho=2.26~$g/cm$^3$, and $d=300~$nm, we obtain
 \begin{equation}
 \omega=3.3 \times 10^5~{\rm rad/s}.
 \end{equation}
 This mechanical frequency is quite high.

  It is possible to estimate also in a simple way the  damping due to air resistance. This factor can actually be important in practice, although it is small in the present case due essentially to the low density of air.   Let us consider the lowest mode again, and evaluate the correction $\omega \rightarrow \omega_1$ because of the air drag. The relevant formula was worked out by Lamb \cite{lamb21}, and is given also in Ref.~\cite{timoshenko55}:
 \begin{equation}
 \omega_1=\frac{\omega}{\sqrt{1+\beta}},
 \end{equation}
 where
 \begin{equation}
 \beta=0.6689 \,\frac{\rho_{\rm air}}{\rho}\frac{a}{d}.
 \end{equation}
 With  $\rho_{\rm air}=1.20~$ kg/m$^3$ and the same parameters as above  we obtain $\beta=1.06\times 10^{-3}$, so that the correction is too small to be observable.

Evidently, also in this time-dependent situation does the rapidly fluctuating Abraham term ${\bf f}^{\rm A}$ not play any role.

\section{Action of the Abraham term in an optical pulse}

We will now give a brief  analysis of how the Abraham force acts on an isotropic medium when an  optical pulse  propagates through it in the $x$ direction.  In this case
 ${\bf f}^{\rm A}$ is longitudinal,
 \begin{equation}
 f^{\rm A}_x= \frac{n^2-1}{c^2}\frac{\partial}{\partial t}{\bf (E\times H)}_x. \label{36}
 \end{equation}
 We focus on the following three aspects:

 \noindent 1) The pulse imparts a mechanical {\it momentum} to the medium; per unit volume it is
 \begin{equation}
 g_x^{\rm mech}=\frac{n^2-1}{c^2}{(\bf{E\times H})}_x. \label{37}
 \end{equation}
 This momentum is due to the forward impulse (kick) given to the particles at the leading edge of the pulse, and a corresponding backward impulse at the trailing edge. In the intermediate region where the pulse can be regarded as a plane wave, the Abraham force fluctuates out.  It was just this accompanying mechanical momentum that was measured in the classic experiments of Jones {\it et al.} \cite{jones54,jones78}.

 \noindent 2) There is a corresponding small {\it displacement} of the particles. For simplicity we assume that the undisturbed particles were at rest. With $N$ denoting the number density of particles, the mechanical momentum received per particle is
 \begin{equation}
 \Delta p= \frac{n^2-1}{Nc^2}{(\bf{E\times H})}_x. \label{38}
 \end{equation}
 The distance $l$ moved by a particle when acted upon by a pulse of duration $\tau$ is thus $l=(\Delta p/m)\tau$, where $m$ is the particle mass. Observe that $l$ is of first order in the small quantity $\Delta p$.

 \noindent 3) Then comes the central point: is there a mechanical {\it kinetic energy}   transformed to the medium? In our opinion the answer is no. The reason is very simple: the kinetic energy per particle is $(\Delta p)^2/(2m)$, thus of second order in $\Delta p$, and hence negligible.

 The conclusion above seems to come at variance with the statement made by Partanen {\it et al.} \cite{partanen17}. These authors presented an impressive numerical calculation of the propagation of an optical pulse in an isotropic medium. Their theoretical analysis contained however an extra term $\delta mc^2$ in the energy expression in the laboratory frame, apparently motivated by relativity, implying that the total energy $E_{MP}$ of the traveling pulse was written in the form $E_{MP}=\hbar \omega +\delta mc^2$. This energy, together with the total propagating photon momentum $p_{MP}$, was taken to transform relativistically through the Lorentz transformation as if $(E_{PM}, p_{MP})$ were the components of a four-vector. This calculation led to $\delta mc^2 =(n^2-1)\hbar \omega,~ p_{MP}=n\hbar \omega/c$  (Eqs.~(2)-(5) in Ref.~\cite{partanen17}), implying in turn that $E_{MP}=n^2\hbar \omega$. However, these values of $E_{MP}$ and $p_{MP}$ do not allow one to use  the Lorentz transformation, as  they are not the energy and momentum components of an energy-momentum tensor whose four-divergence is {\it zero}. They are not the components of a four-vector. This is in  contrast to the properties of  the Minkowski tensor; its vanishing four-divergence  implies that the energy and momentum photon components constitute a four-vector. It is thus necessary to adopt the photon energy in the Minkowski form $\hbar \omega$ (not $n^2\hbar \omega)$, together with the momentum $n\hbar \omega/c$, in the rest frame  in order to use the Lorentz transformation relating energy and momentum in different frames.  (More discussions on this definite restriction on the four-velocity property for the energy and momentum  components   can be found, for instance,  in Refs.~\cite{brevik79} or \cite{moller72}.)

 The correct general-relativistic description of light was pioneered by Gordon \cite{gordon23}, and has been further developed more recently in the research area of transformation optics (see, for instance, Refs.~\cite{leonhardt06} and \cite{leonhardt10}).

 The following remark ought to be added. We neglected above the coupling between forces  exerted by electromagnetic momentum and elastic waves. One might think that such a  coupling is too small to be measurable, but the recent work of Po\v{z}ar et al.\cite{pozar18}  has actually demonstrated experimentally the existence of elastic waves, driven essentially by the momentum of the incident wave. When a laser beam from above impinges upon a horizontal dielectric mirror the radiation pressure launches elastic waves, which spread away from from the source in the lateral direction and carry energy and momentum transferred from the laser pulse. This can in turn give rise to minute displacements of the entrance surface of the dielectric mirror. The detectable amplitudes of the surface were found to be  of order 100 fm.

\section{Conclusion}

The electrodynamics of media is a complicated topic. Our purpose with this note   has been to point out that  care should be taken when interpreting recent experiments and computer experiments. Thus the downward bending of a graphene oxide sheet,  clearly demonstrated experimentally by Kundu {\it et al.} \cite{kundu17} has after all little to do with the Abraham field momentum. And when analyzing the propagation of an optical pulse; cf.  Partanen {\it et al.} \cite{partanen17}, one should observe that there is practically no transfer of a mechanical kinetic  energy to the medium. The mechanical energy transferred   is of second order in the small momentum $\Delta p$ given to the particles   and is thus negligible.

\noindent \section*{Acknowledgements}

\noindent  Thanks go to L. C. Malacarne for alerting me to the experiment \cite{kundu17}, and  go also to  K. S. Hazra for informing me about the actual thickness of the plate. The present  work was supported by the Research Council of Norway, Project 250346.

\end{document}